\documentclass[a4paper,12pt]{article}

\usepackage[utf8]{inputenc}
\usepackage[T1]{fontenc}
\usepackage[a4paper,margin=2.5cm]{geometry}
\usepackage{graphicx}
\usepackage{natbib}
\usepackage[colorlinks=true, linkcolor=blue, citecolor=blue, urlcolor=blue]{hyperref}
\usepackage{amsmath}
\usepackage{booktabs}
\usepackage{authblk}

\usepackage{float}
\usepackage{tabularx}
\usepackage{makecell, ragged2e}
\usepackage[table]{xcolor}
\definecolor{lightgray}{gray}{0.73}
\usepackage{siunitx}
\usepackage{subcaption}
\newcolumntype{Y}{>{\centering\arraybackslash}X}

\newcommand{\tliq}{$T_{\textrm{liq}}$}
\newcommand{\ri}{$n$}
\newcommand{\abbe}{$\nu_d$}

\title{From Patents to Dataset: Scraping for Oxide Glass Compositions and Properties}

\author[1]{Gustavo Laranja Thomaello\textsuperscript{*}}
\author[2]{Thomaz Yeiden Busnardo Aguena\textsuperscript{*}}
\author[3]{Eric Trevelato Costa}
\author[2]{Rafael Baságlia Rosante}
\author[3]{Thiago Rodrigo Ramos}
\author[3]{Daiane Aparecida Zuanetti}
\author[2]{Edgar Dutra Zanotto}

\affil[1]{%
    Department of Chemical Engineering\\
    Federal University of S\~ao Carlos\\
    S\~ao Carlos, S\~ao Paulo, Brazil
}
\affil[2]{%
    Department of Materials Engineering\\
    Federal University of S\~ao Carlos\\
    S\~ao Carlos, S\~ao Paulo, Brazil
}
\affil[3]{%
    Department of Statistics\\
    Federal University of S\~ao Carlos\\
    S\~ao Carlos, S\~ao Paulo, Brazil
}

\date{}

\begin{document}

\maketitle

\def\thefootnote{*}\footnotetext{These authors contributed equally to this work.}

\begin{quotation}
\noindent\textit{\textbf{Abstract}:} 
In this work, we present web scraping techniques to extract in- formation from patent tables, clean and structure them for future use in predictive machine learning models to develop new glasses. We extracted compositions and three properties relevant to the development of new glasses and structured them into a database to be used together with information from other available datasets. We also analyzed the consistency of the information obtained and what it adds to the existing databases. The extracted liquidus temperatures comprise 5,696 compositions; the second subset includes 4,298 refractive indexes and, finally, 1,771 compositions with Abbe numbers. The extraction performed here increases the available information by approx- imately 10.4\% for liquidus temperature, 6.6\% for refractive index, and 4.9\% for Abbe number. The impact extends beyond quantity: the newly extracted data introduce compositions with property values that are more diverse than those in existing databases, thereby expanding the accessible compositional and property space for glass modeling applications. We emphasize that the compositions of the new database contain relatively more titanium, magne- sium, zirconium, niobium, iron, tin, and yttrium oxides than those of the existing bases.

\vspace{9pt} \noindent\textit{\textbf{Keywords}:} web scraping, refractive index, Abbe number, liquidus temperature.
\end{quotation}

\section{Introduction}

Oxide glasses are universal in modern technology -- from domestic and artistic products to bioactive materials for bone regeneration, ionic conductors for batteries, bulletproof windows, optical fibers for efficient communication, precision lenses, smartphone and TV displays, and scientific instrumentation -- owing to their beauty, transparency, durability, and the breadth of tunable properties achievable through composition control. 
However, designing new multicomponent glass compositions with targeted combinations of properties remains challenging. Traditional discovery has relied on empirical rules and iterative trial-and-error, an approach that struggles against the vast, high-dimensional compositional design space and the inherently nonlinear composition–property relationships in multicomponent oxide systems \citep{liu2021machine, montazerian2020model, alcobaca2020explainable, cassar2018predicting, cassar2021predictingB, ravinder2021artificial, gupta2022matscibert, zaki2022extractingA_2021, mannan2023glass}. 
In fact, two decades ago, well before the use of machine learning to understand and develop new glasses, a provocative paper discussed how many glasses could be made by combining the 80 “friendly” elements of the periodic table. The answer was approximately $10^{52}$ \citep{zanotto2004many}. At that time, there was no answer as to how this immense compositional space could be efficiently explored in a reasonable timeframe.

Two properties illustrate the stakes and the difficulty particularly well. The liquidus temperature (\tliq) -- the minimum temperature at which the melt is crystal-free --governs processing windows and susceptibility to devitrification, yet depends sensitively on multioxide interactions. The refractive index (\ri) is a central optical parameter that varies significantly with composition; high-index formulations often come with increased dispersion, quantified by the Abbe number (\abbe), which complicates optical design. Achieving combinations such as high \ri\ with adequately high \abbe\ is nontrivial \citep{cassar2021designingA}. While early neural-network work in glass mapped glass-forming regions rather than predicting properties \citep{tu1996prediction}, the first notable property-prediction study targeted \tliq, employing four datasets, each with up to 850 compositions \citep{dreyfus2003machine}; for \ri, the earliest large-scale, composition-driven models -- already using thousands of compositions -- appeared only in 2020 \citep{ravinder2020deep, bishnoi2021scalable}, as cited by \cite{cassar2021predictingB}; and the first extensive treatment involving the Abbe number, \abbe, likewise came with \cite{cassar2021predictingB}, who also consolidated the largest datasets available at that time ($\approx 33{,}000$ for \tliq; $\approx 45{,}000$ for \ri; and $\approx 22{,}000$ \abbe). Accurately predicting \tliq, \ri, and \abbe\ directly from composition is therefore both scientifically and economically significant, motivating the integration of data-driven modeling to complement, and sometimes replace, slow empirical iteration \citep{cassar2018predicting, cassar2021designingA, cassar2021predictingB, alcobaca2020explainable, liu2021machine, mastelini2022machine, mannan2023glass}.

In this context, artificial intelligence, particularly machine learning (ML), offers a transformative opportunity by leveraging existing datasets to develop data-driven models that can expedite the discovery of new glass compositions with desired combinations of properties. Effective ML methods for glass property prediction rely on access to large, structured, and up-to-date datasets. Historically, two proprietary compilations -- SciGlass and INTERGLAD -- aggregated hundreds of thousands of glass compositions and their corresponding properties, and have underpinned many ML studies. These resources have demonstrated that data-driven models can achieve uncertainties comparable to experimental scatter for several properties, including \tliq, refractive index, and Abbe number \citep{alcobaca2020explainable, cassar2018predicting, cassar2021designingA, cassar2021predictingB, ravinder2021artificial, venugopal2021looking, zaki2022extractingA_2021, mastelini2022machine, hira2024reconstructingA, hira2025matskraft, mannan2023glass}.

However, both databases have clear limitations: they either require paid access, rely on manual curation, or are slowly updated. SciGlass, in particular, has historically depended on manual extraction and has been discontinued since 2014, while INTERGLAD is closed-access and similarly curated \citep{venugopal2021looking, zaki2022extractingA_2021}. As a result, recent and rich sources of information are dispersed across the literature and industry documents rather than being centralized in a single, living resource \citep{venugopal2021looking}. Building on this problem, here we propose an alternative approach: creating an ML-ready dataset by systematically mining patents and integrating oxide compositions with their reported \tliq, \ri, and \abbe\ values.

The glass research community has begun to address this bottleneck using natural language processing (NLP) to extract materials data from research articles \citep{venugopal2021looking, zaki2022extractingA_2021, zaki2022naturalB, gupta2022matscibert, hira2024reconstructingA, sheth2024extraction, hira2025matskraft}. Significant advances include domain-adapted language models\footnote{Pre-trained models adjusted to excel in a specific field, such as finance or healthcare, by incorporating domain-specific data and terminology, thereby improving performance on tasks within that specialized area.} and table-aware extractors that convert text and tables into structured records suitable for ML \citep{gupta2022matscibert, gupta2022discomat_2023, hira2024reconstructingA, sheth2024extraction, hira2025matskraft}. Yet, all existing NLP pipelines in materials science mine journal articles. Patents -- which frequently report exhaustive tables of compositions and measured properties for novel glasses -- remain largely untapped as a systematic data source \citep{hira2024reconstructingA}. Given the pace and volume of patenting in optics, displays, and specialty glass, this represents a substantial, underexploited reservoir of structured examples ideally suited for ML-ready databases. 

Beyond data collection, previous extraction efforts have rarely conducted formal significance testing to quantify the incremental value of newly harvested composition–property pairs -- particularly against legacy resources such as SciGlass and INTERGLAD -- leaving the actual data gain in ML predictions unmeasured \citep{venugopal2021looking, gupta2022matscibert, hira2024largeB, sheth2024extraction}. An exception is the recent article by \cite{hira2025matskraft}, which compared the total gain of obtained compositions relative to these databases and included range plots for 18 different glass properties.

In this article, we address this gap by developing an automated web scraping pipeline targeted at Google Patents to assemble a comprehensive dataset of oxide glass compositions initially annotated with liquidus temperature, refractive index, and Abbe number. Web scraping enables scalable, automated collection of semi-structured information from the web; as a research technique, it can rapidly produce datasets far beyond the practical reach of manual collection \citep{landers2016primer, khder2021web} and is now standard practice in data-centric workflows \citep{rundeldogucu2021web, khder2021web}. Our approach tailors scraping to the key sections of patent text (e.g., claims, examples, tabulated data) and applies regular expressions with post-processing for chemical named-entity recognition, unit parsing, and linkage of oxide composition (mol\%, wt\%) to reported \tliq, \ri, and \abbe. In addition to harvesting, we integrate the patent-scraping-derived dataset with entries from INTERGLAD and SciGlass and perform statistical analyses to assess the incremental information contributed by our approach to the overall glass composition - property dataset. Specifically, we compare distributions and statistics of key properties across datasets to evaluate how the newly scraped records expand the known compositional and property spaces. This analysis provides a quantitative view of the added coverage for \tliq, \ri, and \abbe.

This patent-centric strategy offers four advantages over reliance on legacy databases or article-only NLP: applicability (patents reflect the latest industrial innovations), density (single documents often include dozens of composition–property examples), scalability (continuous harvesting avoids the stagnation typical of manually curated archives), and precision (when well extracted, the data obtained are accurate and do not depend on the performance of NLP predictive models). Together, these features promise a living, extensible dataset that can foster contemporary glass research and development \citep{venugopal2021looking, ravinder2020deep, zaki2022extractingA_2021, gupta2022matscibert, hira2024largeB, sheth2024extraction, hira2025matskraft}.

Therefore, our goals are: i) to construct a scraped and cleaned dataset from patents that show oxide glass compositions and their respective liquidus temperature, refractive index, and Abbe number, and ii) to demonstrate that such a dataset is structured, reproducible, and suitable for future ML applications in accelerated glass design. This resource lays the groundwork for subsequent studies in which predictive and inverse-design models may be developed to obtain glassy compositions with desired properties \citep{cassar2021predictingB}.

This manuscript introduces the methodology for creating the curated dataset and illustrates its characteristics through descriptive and statistical analyses. The introduction provides a brief review of data-driven glass science and information extraction in materials, positioning patent scraping within prior literature. Section~\ref{sec:methods} details the patent-scraping pipeline, including query design, parsing and normalization of compositions and properties, and data cleaning. Section~\ref{sec:results} characterizes the resulting dataset -- covering composition ranges, property distributions, and illustrative insights -- and reports analyses that quantify the incremental coverage obtained when augmenting INTERGLAD and SciGlass with patent-scraped data. Finally, Section~\ref{sec:discussion} presents concluding remarks and outlines directions for future work.

\section{Methodology}\label{sec:methods}

Web scraping is the automated and systematic extraction of information from the web, followed by the transformation of that content into structured datasets suitable for analysis, in contrast to manual data collection. Operationally, web scraping involves three main stages: (i) locating and retrieving relevant web pages (URLs), (ii) extracting the desired data elements, and (iii) transforming unstructured outputs into structured formats for storage and downstream use (e.g., CSV tables).

Within this framework, the extraction workflow can be understood as comprising two ordered components: the \textit{crawler}, which enumerates and queues pages to be systematically fetched, and the \textit{scraper}, which parses the stored files (typically HTML or XML) to capture the target fields. When the scope is broad, crawling demands high request rates, storage, and queue coordination, which may overload specific domains. Search robots such as Googlebot exemplify this pattern of systematic scanning \citep{landers2016primer, rundeldogucu2021web, khder2021web}.

In the present work, web crawling was manually controlled through predefined lists, without automatic link discovery. A fixed set of Google Patents URLs was read from a file and processed using Selenium \citep{selenium2025}. For each URL, the crawler waits for the page to load and extracts the target tables -- specifically, those containing composition and glass property data, hereafter referred to as \texttt{patent\_tables}. The processing queue is deduplicated by checking for existing JSON outputs for each patent and managed through control lists that track whether the corresponding tables are present or absent.

After loading, the scraper collects standardized metadata exposed in meta tags (e.g., inventors, assignee, dates, PDF link) and captures the raw HTML of the patent table sections. It then applies lightweight heuristics to label the composition unit detected in the tables (molar, mass-based, or uncertain) using regular expressions and normalized keywords. Each patent is serialized into a single JSON file that retains the URL, metadata, and the HTML table, ensuring traceability across all subsequent processing stages. 

All scripts used for the scraping workflow and preprocessing steps are available at
\url{https://github.com/thiagorr162/glass_patents}\footnote{This repository will be publicly available after manuscript publication.}.

\subsection{Source Identification: Google Patents}

The quality, coverage, and reproducibility of a web scraping pipeline depend directly on the choice of source. Beyond recognizing the central role of data for decision-making and research, it is advisable to develop a “data source theory” before collecting and to reflect on: why the source exists, what types of information it offers, how the data therein are organized and operationalized, and whether they can answer the study’s questions; this includes inspecting the raw HTML, mapping variables, and defining how they will be extracted \citep{landers2016primer}. Because scraping relies heavily on source selection, it is necessary to prioritize reliable, stable, and preferably publicly accessible portals to ensure quality and reduce the burden of post-extraction curation. In summary, data are valuable, but they become useful for decision-making when their origin is clear and technically fit for purpose \citep{khder2021web}.

In light of this framing, we chose Google Patents as the primary source because it aggregates standardized, examined legal documents that are rich in metadata and organized as HTML, which facilitates structure mapping and deterministic extraction in line with the ``data source theory''. Moreover, in the reference and recent literature on data extraction applied to materials science, extraction is performed mostly via a common stack of APIs and publisher portals (Crossref/Elsevier/ScienceDirect) \citep{venugopal2021looking, gupta2022matscibert, zaki2022naturalB, sheth2024extraction, hira2024largeB}. By contrast, although patents are listed as reliable sources for extraction \citep{ravinder2020deep, hira2024largeB}, we did not identify reports of extracting compositions and materials properties from patents. Thus, anchoring the scraping in Google Patents introduces an innovative element relative to existing article-centered pipelines, without sacrificing robustness and reproducibility, which are decisive attributes for dataset quality.

\subsection{Crawler and Scraper Implementation in Python}

The developed architecture was divided into two main modules -- a crawler and a scraper -- which work together in a coordinated manner. The crawler focuses on page navigation and raw collection, whereas the scraper targets the specific extraction of information.
In our context, the crawler automatically traversed the target patent URLs, managed the queue of pages to visit, and retrieved the relevant HTML elements from each patent. We first compiled a list of patents relevant to the topic by searching Google Patents with a set of keywords and multiple combinations thereof. The results were exported, deduplicated, and consolidated into a URL list consumed by the crawler. The crawler visited each Google Patents URL, simulating a user session by employing the Selenium library via the Mozilla Firefox web browser.

Running in parallel, the scraping module was responsible for extracting structured data from the loaded HTML pages. While the crawler actually loaded each URL, the scraper searched header elements using a normalized list of oxide names. In this manner, patents containing tables deemed relevant were separated and, when present, the table segments were serialized along with other metadata (title, type, application/publication numbers, inventors, etc.) into per-URL JSON files, as shown in Figure \ref{fig:json_example}. We also defined a stable identifier, referred to as \texttt{patent\_id}, for each extracted table block, in the format \texttt{<publication\_number>\_block\_<k>}. The prefix \texttt{<publication\_number>} encodes the patent URL, and the suffix \texttt{<k>} differentiates multiple tables within the same patent.

\begin{figure}[H]
\footnotesize
\centering
\begin{minipage}{0.9\linewidth}
\begin{verbatim}
{
  "url": "https://patents.google.com/patent/{AUTH}XXXXXXX/en",
  "title": "Patent Title Example",
  "type": "patent",
  "description": "Short summary of the patent.",
  "application_number": "{AUTH}:XX/XXX,XXX",
  "publication_number": "{AUTH}XXXXXXX",
  "pdf_url": "https://patentimages.storage.googleapis.com/.../{AUTH}XXXXXXX.pdf",
  "inventors": [
    "First Inventor",
    "Second Inventor"
  ],
  "assignee": "Company Name",
  "date": [
    "Application Date",
    "Publication Date"
  ],
  "html_tables": [
    "<patent-tables> ... </patent-tables>"
  ]
}
\end{verbatim}
\end{minipage}
\caption{Example structure of a JSON file generated by the scraping module, showing the standardized metadata fields and raw table content extracted from Google Patents.}
\label{fig:json_example}
\end{figure}

This identifier accompanies all CSVs and subsequent steps, serving as a traceability key to recover the metadata from the associated JSON and, when needed, reconstruct the publication number to compose the Google Patents link. At this stage, we applied lightweight content heuristics: we searched for occurrences of unit indicators in the table HTML (terms containing ``mol'' or mass variants such as ``wt\%'', ``weight \%'', ``mass\%''). Unit characterization was performed at the document (per-patent) level: we aggregated evidence across all tables in the table list and assigned units. The identified unit could be \textit{mol}, \textit{mass}, \textit{both}, or \textit{none}. This label only guides unit adjustments and mass$\leftrightarrow$mol conversions in later steps, without altering values at this stage. We also adopted basic text normalization (removal of diacritics, space normalization, and case standardization) to facilitate later searches. Importantly, at this stage, we did not perform cell-by-cell cleaning or explicit transformation of chemical subscripts. The saved snippets served as input for subsequent filtering.

Filtering for relevance by oxide compounds of interest (such as SiO$_2$, Al$_2$O$_3$, Na$_2$O, etc.) therefore occurred only in the subsequent stage: the serialized \texttt{patent\_tables} fragments were revisited, and only those in which at least one of these compounds was detected were preserved in an organized repository for later processing. Next, the tables were segmented into blocks (obtained by splitting the table HTML) and converted into tabular structures with the support of header heuristics: (i) identification of the start row when there were $\geq 2$ desired compounds; (ii) verification that the header contained property keywords (e.g., ``refractive,'' ``Abbe,'' ``liquidus,'' ``CTE,'' ``nd''); (iii) technical limits for header width and number of columns, aimed at discarding noisy layouts. The approved tables were then materialized as CSV files per block, with a source identifier assigned to each URL/block, preserving the content for subsequent analyses, as illustrated in Table \ref{tab:example_patent_table}.

\begin{table}[H]
\centering
\caption{Example of a consolidated table extracted from a patent, showing oxide compositions (mol.\%), measured properties, and the corresponding \texttt{patent\_id}.}
\label{tab:example_patent_table}
\footnotesize
\begin{tabular}{ccccccccccc}
\toprule
Nb$_2$O$_5$ & P$_2$O$_5$ & Na$_2$O & \dots & TiO$_2$ & K$_2$O & SrO & \ri & \tliq (°C) & \abbe & \texttt{patent\_id} \\
\midrule
35.9 & 22.1 & 1.6 & \dots & 15.0 & 5.0 & 2.1 & 1.950 & 673 & 18.6 & \texttt{us11485676b2\_b12} \\
38.8 & 20.9 & 0.3 & \dots & 15.0 & 15.4 & 0.9 & 1.984 & 689 & 19.2 & \texttt{us11485676b2\_b12} \\
37.9 & 21.4 & 0.0 & \dots & 15.0 & 12.1 & 1.5 & 1.967 & 702 & 19.0 & \texttt{us11485676b2\_b12} \\
\bottomrule
\end{tabular}
\end{table}

In this way, the crawler + scraper architecture allowed us to comprehensively traverse the set of patents (via the crawler) and then isolate and structure only the desired information (via the scraper), with an intermediate phase of \texttt{patent\_tables} + metadata serialization and a subsequent phase of filtering and tabulation based on header/compound heuristics. This dual approach ensures efficiency and modularity: similar to what is described by \cite{khder2021web}, web crawling navigated through all relevant URLs, while web scraping “selectively retrieved the specified data” from each page for use in later analyses. 

\subsection{Merging Extracted Tables}

Following the consolidation of the individualized tables, a consolidation by parts was carried out for the files generated in the previous phase. The objective was to unify the exported valid blocks (those that passed the pre-filtering stage described) under a single schema and materialize them in a consolidated file. To this end, the script recursively traversed the input CSVs and, before any merge, detected the headers actually present: by reading only the header line of each file (without loading data), an ordered union of the column names was formed, preserving the first occurrence of each label and logging warnings when any file’s header could not be read. This union defined the target schema to be imposed on the result.

With the unified schema established, incremental writing of the consolidated file was initiated. First, the final header was written only once. Next, each source CSV was read filtering only the columns of the target schema and then reindexed to impose the same column order and insert missing ones (filled with \texttt{NaN}). Empty rows were discarded, and the remaining rows were appended to the output file without rewriting the header, in a vertical process (with no key-based join or row deduplication).

Finally, simple post-processing was performed on the consolidated file: the final CSV was reopened and all 100\% empty columns (i.e., those that remained \texttt{NaN} in all rows) were removed. The result was rewritten in UTF-8 and the output path reported at the end of the process. If no input files were found or no valid header was detected, execution was terminated with explanatory messages, maintaining transparency regarding the state of the consolidation. This initial and unstructured dataset was saved and, from it, successive filtering procedures were carried out to construct the \ri, \abbe, and \tliq\ datasets.

\subsection{Dataset Cleaning and Revision}

After extraction and preliminary consolidation, the resulting dataset still contained redundant entries, structural inconsistencies, and residual text fragments inherited from heterogeneous patent tables. To transform this raw material into reliable, analyzable data, a structured cleaning workflow was applied—beginning with an initial filtering stage that enforces consistency, relevance, and traceability.

\subsubsection{Initial Filtering}

The initial filtering step reads the previously consolidated dataset in batches, pre-selecting only the columns considered relevant to the process -- the source identifier (\texttt{patent\_id}), composition columns (obtained from the list of desired compounds), and property columns (detected by a broad set of keywords and patterns). It is important to mention that reading in chunks and incremental writing via \texttt{append} were adopted to enable processing a volume on the order of tens of gigabytes, thus avoiding the memory costs associated with \texttt{pandas}' traditional \texttt{concat} method \citep{reback2020pandas}.

For each batch, a simple numeric normalization is performed: hyphens
-- commonly used to represent zero or unmeasured properties in patents -- are replaced with zeros, values are coerced to numeric, and missing values are filled with zeros. Next, the filter isolates the subset of compositions (only oxide columns), removes any columns derived from the sum of compounds (labels containing ``+'' alongside compound names, e.g., SiO$_2$+B$_2$O$_3$+Al$_2$O$_3$), and applies the closure criterion ($\sum \text{composition} \approx 100$ with a tolerance of 0.5) to retain only rows with valid compositions. In parallel, the filter selects the subset of properties, keeping only rows with effective content (row-wise sum different from zero, i.e., at least one non-null property entry). The two views -- composition and properties -- are then intersected at the row level, ensuring that each retained record simultaneously has a closed composition and at least one reported property.

To preserve essential metadata, the \texttt{patent\_id} is retained at the end of each batch. The unit declared in the source document metadata (\textit{mol}, \textit{mass}, \textit{both}, or \textit{none}) during extraction is retrieved at this stage through a mapping constructed from the \texttt{patent\_id} itself. Each filtered batch is stored as an independent CSV file containing the ``parts,'' making the process fault-tolerant and scalable.

In the final phase, these filtered parts are combined by the union of headers and vertical append, with column realignment and filling of missing values with zero, resulting in a new dataframe. Finally, columns that are entirely null (100\% zeros after alignment) are removed, preserving a compact table containing only (i) closed compositions, (ii) properties identified by broad keywords, and (iii) the \texttt{patent\_id} and unit metadata required for subsequent analysis steps.

\subsubsection{Property-Specific Filtering and Consolidation}

Starting from the dataframe generated in the initial filtering stage, we proceeded toward specific filtering to isolate, by target property, subsets containing (i) a valid composition and (ii) the standardized property, preserving traceability via \texttt{patent\_id} and unit. In all cases -- refractive index (\ri), Abbe number (\abbe), and liquidus temperature (\tliq) -- the procedure reads the input in chunks and writes results via \texttt{append}, ensuring scalability for large volumes of data. At the end of the pipeline, we also emit a patent contributions report that, for each property column, lists the patents that actually populated it (values $\neq 0$). 

Because the capture yields a highly heterogeneous set of column labels, each candidate column retains the tags of the patents that contributed to it. We then partition these columns into three classes: (i) heterogeneous columns with an explicit unit or wavelength; (ii) ``generic'' columns lacking such declarations; and (iii) false positives. A dictionary drives resolution across both routes: it normalizes labels into standardized targets (e.g., \ri($\lambda$); \tliq\ by unit) and holds blacklists to suppress class (iii). Its primary role, however, is to encode a manually curated disambiguation map -- lists of \texttt{patent\_ID}s $\rightarrow$ standardized column -- 
built by domain experts. 

For class (i), values are simply reallocated to the appropriate standardized column (e.g., ``liquidus temperature (°C)'', ``Tliquidus °C'', ``liq. C'' $\rightarrow$ \tliq\ (°C)). For class (ii), curators determine, patent by patent, the missing unit or wavelength from the source tables and then map those patents to the correct standardized column (e.g., \{\texttt{patent1}, \texttt{patent2}, \texttt{patent3}\} $\rightarrow$ \tliq\ (°C)). Columns in class (iii), together with their associated compositions, are discarded.

\subsubsection{Refractive Index and Abbe Number Processing}

In the specific filter for the refractive index (\ri), the dataframe is reduced to oxide columns (composition) and plausible \ri\ columns, discarding as non-\ri\ those whose domain contains values outside the range $(1, 5]$ -- that is, adopting a broader interval than typical \ri\ values, which usually lie between 1 and 1.5. This broad filter, however, also allowed columns related to other properties (e.g., density) to pass through. To address this, a blacklist of columns unrelated to \ri/\abbe\ was manually curated. Columns explicitly mentioning \abbe\ are preserved for later consolidation. 

At the row level, a ``Merged Refractive Index'' marker is calculated: if there is exactly one refractive index value, it is retained; if there is more than one, the row is marked as ambiguous ($-1$) -- typically corresponding to cases where \ri\ was measured at different wavelengths -- and removed from the simple \ri\ subset. The resulting dataframe contains valid composition, \ri, \abbe, and \texttt{patent\_id}/unit information, accompanied by a contributions-by-patent CSV file.

The consolidation of units for \ri\ corresponds to normalization by wavelength ($\lambda$). A JSON dictionary is used to map heterogeneous labels to standardized targets: \texttt{nD} (589.3\,nm), \texttt{nG} (435.8\,nm), \texttt{nF} (486.13\,nm), \texttt{nH} (404.7\,nm), and \texttt{nC} (656.3\,nm). Subsequently, the heterogeneity of declared labels is resolved into these standardized columns through manual mapping and expert curation, yielding a single declared value for each wavelength. In general, only one wavelength is declared per composition, typically corresponding to \texttt{nD}. 

Generic columns (those without explicit $\lambda$) are resolved by a manually curated patent-to-wavelength map, also stored in the JSON dictionary. When a row contains a single generic value and the corresponding patent belongs to a specific $\lambda$ list, the value is assigned to the appropriate standardized column. In the end, such generic columns are discarded. The \abbe\ columns are gathered under a single standardized ``Abbe Number'' field. Rows lacking any standardized \ri\ value are excluded. 

The final output of this stage is a dataframe standardized for cross-patent and cross-property comparisons.

\subsubsection{Liquidus Temperature Processing}

In the specific filter for liquidus temperature (\tliq), the selection of property columns is guided by regular expressions applied to column names, using patterns designed to capture different tags such as ``liquidus,'' ``tliq,'' and ``tl.'' Due to the scale of the input data, writing is performed iteratively in parts at this stage. These parts are subsequently unified through a ``merge parts'' procedure with header alignment. The intermediate result consists of a dataframe along with the corresponding contributions-by-patent CSV file.

The consolidation of units for \tliq\ normalizes measurements expressed in °C, °F, and K, and, when applicable, accounts for the experimental measurement condition (Internal, Air, or Platinum) according to a JSON-based mapping. Columns mapped with certainty are grouped to form standardized targets -- for example, \tliq\ (°C), \tliq\ Air (°C), \tliq\ Platinum (°C), \tliq\ (°F), and \tliq\ (K) -- and the original heterogeneous columns are then removed. 

Generic columns (those without explicit units) are resolved using manually curated patent-to-unit lists, analogous to the procedure adopted for \ri. An additional filtering step enforces plausibility by range (e.g., 450–1900 °C) to retain only a single valid property candidate per composition. This last step addresses specific cases where liquidus temperature and viscosity were measured at the same temperature. 

The final output of this stage is a dataframe containing compositions with standardized \tliq\ values and complete traceability metadata.

\subsection{Compositional Basis Adjustment and Conversion}

As a final step, we executed the adjustment and normalization of the compositional basis in both property sets (\ri/\abbe\ and \tliq). The objective was to resolve cases previously labeled as uncertain in the unit field (values \textit{both} or \textit{none}) and, subsequently, to produce equivalent versions of the compositions on both molar and mass bases. The procedure starts from the dataframes already consolidated by property and operates only on metadata and composition columns, preserving the \texttt{patent\_id} and all previously standardized property content.

The unit adjustment follows the same rationale for both properties: identify rows with uncertain units and reassign the correct basis (\textit{mol} or \textit{mass}) using manually curated patent lists stored in the same JSON dictionary used in the properties stage. For each entry, the patent is extracted from the \texttt{patent\_id} (by parsing its uppercase suffix) and compared with the curated sets: if the glass composition is identified as molar, the unit is set to \textit{mol}; if it corresponds to weight percentage, it is set to \textit{mass}. Rows that remain uncertain are recorded in an auxiliary \texttt{.txt} file in the format ``patent link $\rightarrow$ uncertain label'', serving as a queue for manual inspection and process auditing. As output, we obtain dataframes whose compositions are fully defined with respect to their molar or mass basis.

Next, the conversion of the compositional basis is performed through two mirrored conversion routines: (i) mass$\rightarrow$mol and (ii) mol$\rightarrow$mass. Both receive the unit-adjusted dataset and a dictionary of compounds with their respective molar masses. For each row whose unit differs from the target, the composition columns are selected and the conversion is computed row by row by division or multiplication by the molar masses, followed by normalization to sum to 100 (rounded to two decimal places). Rows already on the target basis are preserved without redundant copying. By design, the unit column is removed in the converted dataset (since the basis becomes intrinsic to the generated file), and all other fields (e.g., standardized \ri, \abbe, and \tliq) remain unchanged. 

The result is two complementary datasets per property flow: one on a molar basis and another on a mass basis, suitable analyses that require a strict comparison between compositions expressed under different conventions.

\begin{table}[H]
\centering
\caption{Excerpt from the cleaned dataset showing oxide compositions (mol\%), the refractive index (\ri D, where \ri D represents the refractive index measured at the sodium line, but others are available), and the Abbe number (\abbe).}
\label{tab:example_dataset}
\footnotesize
\begin{tabular}{ccccccccccc}
\hline
SiO$_2$ & P$_2$O$_5$ & ZrO$_2$ & Na$_2$O & Al$_2$O$_3$ & CaO & K$_2$O & B$_2$O$_3$ & ... & \ri D & \abbe \\
\hline
12.32 & 0.00 & 3.23 & 0.00 & 0.00 & 0.00 & 0.00 & 29.72 & ... & 1.8046 & 40.6 \\
12.15 & 0.00 & 2.96 & 0.00 & 0.00 & 0.00 & 0.00 & 28.02 & ... & 1.8082 & 40.4 \\
6.26 & 0.00 & 3.59 & 0.00 & 0.00 & 0.00 & 0.00 & 35.18 & ... & 1.8107 & 40.6 \\
6.26 & 0.00 & 3.60 & 0.00 & 0.00 & 0.00 & 0.00 & 35.18 & ... & 1.8093 & 41 \\
6.27 & 0.00 & 2.83 & 0.00 & 0.00 & 0.00 & 0.00 & 35.32 & ... & 1.8118 & 40.2 \\
$\vdots$ & $\vdots$ & $\vdots$ & $\vdots$ & $\vdots$ & $\vdots$ & $\vdots$ & $\ddots$ & $\vdots$ & $\vdots$ & $\vdots$ \\
65.05 & 2.23 & 0.00 & 0.00 & 0.62 & 0.00 & 0.67 & 0.00 & ... & 1.5586 & -- \\
\hline
\end{tabular}
\end{table}

\section{Results}\label{sec:results}

This section presents the main outcomes of the data extraction and analysis pipeline. First, we describe the construction and characterization of the Patents database, including the number of compositions retrieved and their correspondence with existing glass property repositories. We then analyze the novelty and diversity of the collected data in terms of both properties and chemical compositions, highlighting how the extracted information complements established datasets such as SciGlass and INTERGLAD.

\subsection{Patents Database Overview}


The data extraction process resulted in a database containing \textbf{9,432 glass compositions}, obtained through the integration of two complementary subsets. The first subset, extracted using multiple combinations of keywords related to \textit{glasses} and \textit{liquidus temperature}, comprises \textbf{5,696 compositions}, with all temperature values reported in degrees Celsius. The second subset, generated from keyword combinations associated with \textit{glasses} and \textit{refractive index}, includes \textbf{4,298 compositions}, with all refractive index values measured at the sodium D-line, and an example of these refractivity data is shown in Table~\ref{tab:example_dataset}. Notably, all \textbf{1,771 compositions} with reported Abbe numbers were retrieved as a byproduct of the refractive index search. Because of the strong correlation between these two optical properties, many patents reporting refractive index values also include corresponding Abbe number measurements. Consequently, no dedicated extraction was performed using keyword combinations specific to the Abbe number.

The compiled data were then analyzed to quantify their coverage and to assess how the extracted compositions and property values compare with existing glass databases.
To contextualize the scope of the compiled data, the Patents database developed in this work was compared with two established sources in glass science: SciGlass (abbreviated as SG) and INTERGLAD (abbreviated as INTG). The comparison focused on three key properties -- liquidus temperature, refractive index, and Abbe number -- and aimed to quantify both the extent and originality of the information extracted from patents relative to these reference databases.


Some compositions appear simultaneously in more than one of the considered databases. To avoid biased interpretations, the Patents data were analyzed both in full and after removing entries already present in other sources. Duplicates were identified by comparing oxide compositions with identical constituent proportions used as the criterion for equivalence. 


Accordingly, the subsets labeled Patents-INTG, Patents–SG, and Patents–Unique correspond, respectively, to compositions not found in INTERGLAD, in SciGlass, and in either of the two databases. Percentages in parentheses indicate the relative contribution of each subset compared with the combined total number of compositions in SciGlass+INTERGLAD (deduplicated).

\begin{table}[H]
\centering
\footnotesize
\caption{Number of compositions across different databases. 
SG = SciGlass, INTG = INTERGLAD. Patents–INTG and Patents–SG denote compositions not present in the corresponding databases.}
\label{tab:dataset_comparison}
\renewcommand{\arraystretch}{1.15}
\begin{tabular}{lcccc}
\toprule
\textbf{Source} & \textbf{Total} & \textbf{Liquidus Temperature} & \textbf{Refractive Index} & \textbf{Abbe Number} \\
\midrule
SG+INTG & 82,898 (100\%) & 37,870 (100\%) & 55,479 (100\%) & 27,150 (100\%) \\
SG & 69,216 (83.5\%) & 32,769 (86.5\%) & 45,364 (81.8\%) & 22,249 (82.0\%) \\
INTG & 20,197 (24.4\%) & 7,114 (18.8\%) & 15,337 (27.6\%) & 6,493 (23.9\%) \\
Patents & 9,423 (11.4\%) & 5,968 (15.8\%) & 4,298 (7.7\%) & 1,771 (6.5\%) \\
Patents–INTG & 8,872 (10.7\%) & 5,253 (13.9\%) & 4,123 (7.4\%) & 1,682 (6.2\%) \\
Patents–SG & 7,372 (8.9\%) & 4,041 (10.7\%) & 3,715 (6.8\%) & 1,358 (5.0\%) \\
Patents–Unique & 7,234 (8.7\%) & 3,938 (10.4\%) & 3,673 (6.6\%) & 1,334 (4.9\%) \\
\bottomrule
\end{tabular}
\end{table}


Based on the data in Table~\ref{tab:dataset_comparison}, the extraction performed in this study increases the available information by approximately 10.4\% for liquidus temperature, 6.6\% for refractive index, and 4.9\% for Abbe number. Although these percentages may seem modest, the impact extends beyond quantity. The newly extracted data introduce compositions with property values that are more diverse than those in existing databases, thereby expanding the accessible compositional and property space for modeling applications.

Another motivation for developing a new data source lies in the limitations of existing glass property repositories. In particular, while the SciGlass database remains the largest and most comprehensive compilation of oxide glass compositions and properties to date, it has not been updated since its discontinuation in 2014. Consequently, a large fraction of glass compositions published or patented after this period are not represented in that resource. 

\begin{figure}[H]
    \centering
    \includegraphics[width=0.65\textwidth]{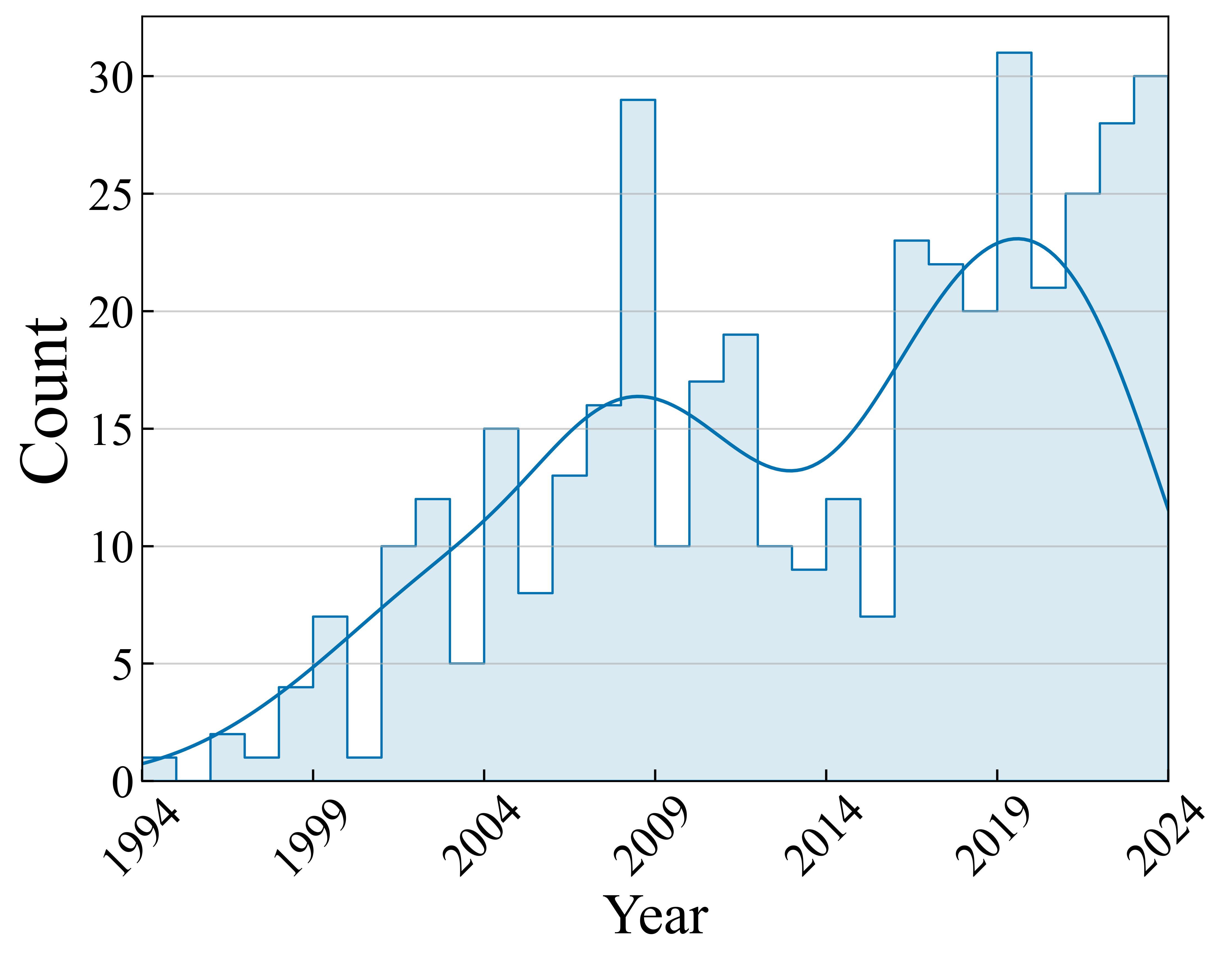}
    \caption{Number of patents included in the dataset per year (data-contributing only).}
    \label{fig:patent_years}
\end{figure}

To illustrate another important aspect of the information extracted from patents, Figure~\ref{fig:patent_years} shows the distribution of patent publication years from which this dataset was extracted. The steadily increasing number of patents (each of which can contribute dozens or even hundreds of distinct compositions), especially after 2014, clearly indicates that most of the compositions collected through our scraping of Google Patents correspond to developments that emerged after SciGlass ceased to be maintained. This reinforces the notion that the new dataset captures information that is both contemporary and largely absent from SciGlass.

\subsection{Property and Compositional Novelty}

The primary objective of this work is the development of a database designed to complement established repositories such as SciGlass and INTERGLAD, thereby enhancing the performance of predictive models trained with data from these sources. Within this context, it is also essential to assess the degree of originality of the collected compositions. This comparative analysis enables the identification of how effectively the patent-derived data expand the compositional and property space already covered by existing databases, providing a more accurate estimate of the potential impact of the developed dataset when integrated with other repositories.

\begin{figure}[H]
\centering

\begin{subfigure}[t]{0.4855\textwidth}
\centering
\includegraphics[width=\linewidth]{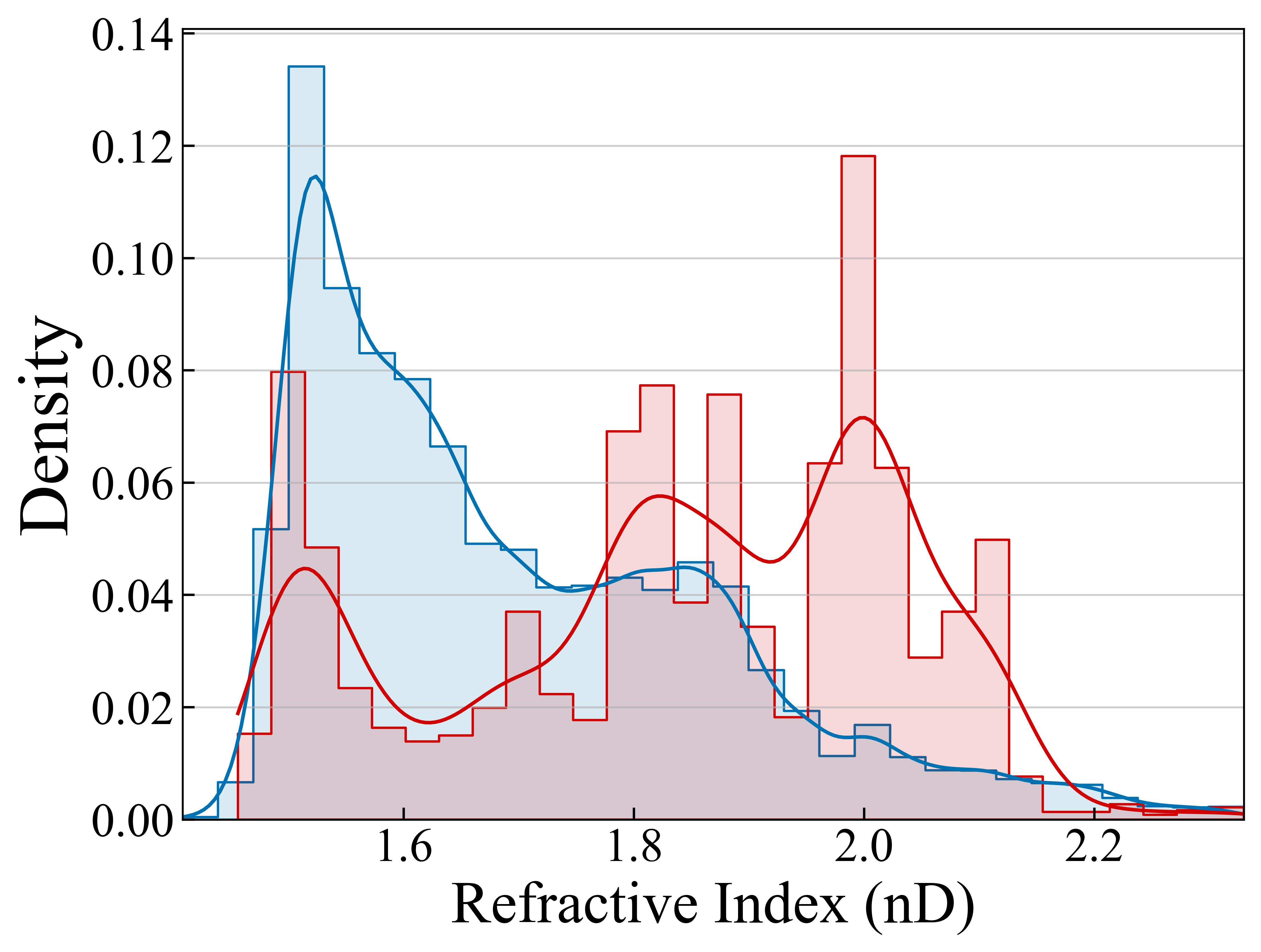}
\caption{}
\label{fig:rihist}
\end{subfigure}
\hfill
\begin{subfigure}[t]{0.499\textwidth}
\centering
\includegraphics[width=\linewidth]{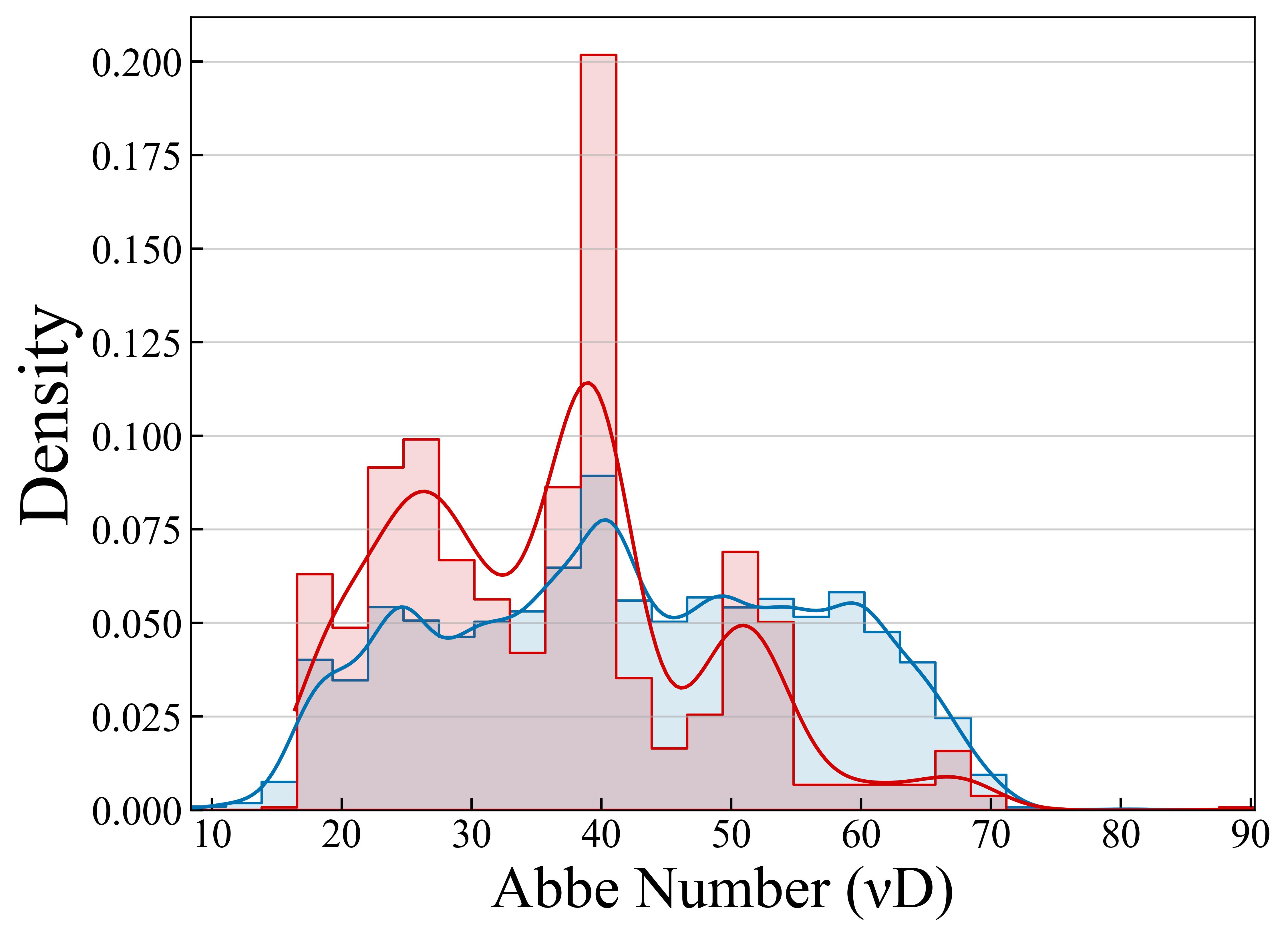}
\caption{}
\label{fig:abbehist}
\end{subfigure}

\vspace{0.5em}

\vspace{-2.4em}
\begin{minipage}{0.3\textwidth}
\centering
{
\includegraphics[width=\linewidth]{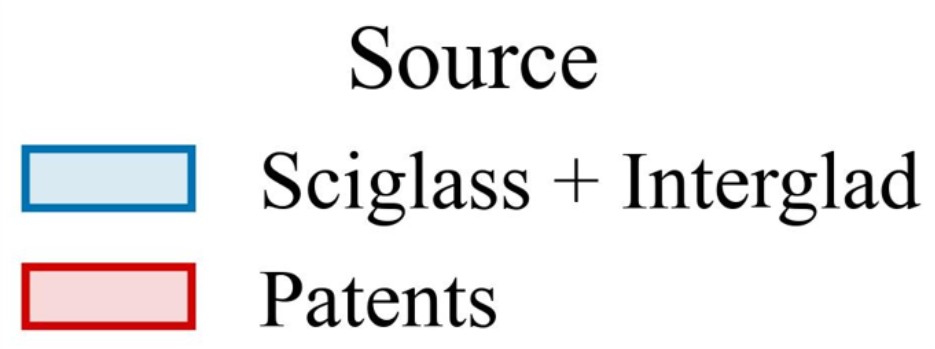}
}
\end{minipage}

\vspace{0.5em}

\begin{subfigure}[t]{0.495\textwidth}
\centering
\includegraphics[width=\linewidth]{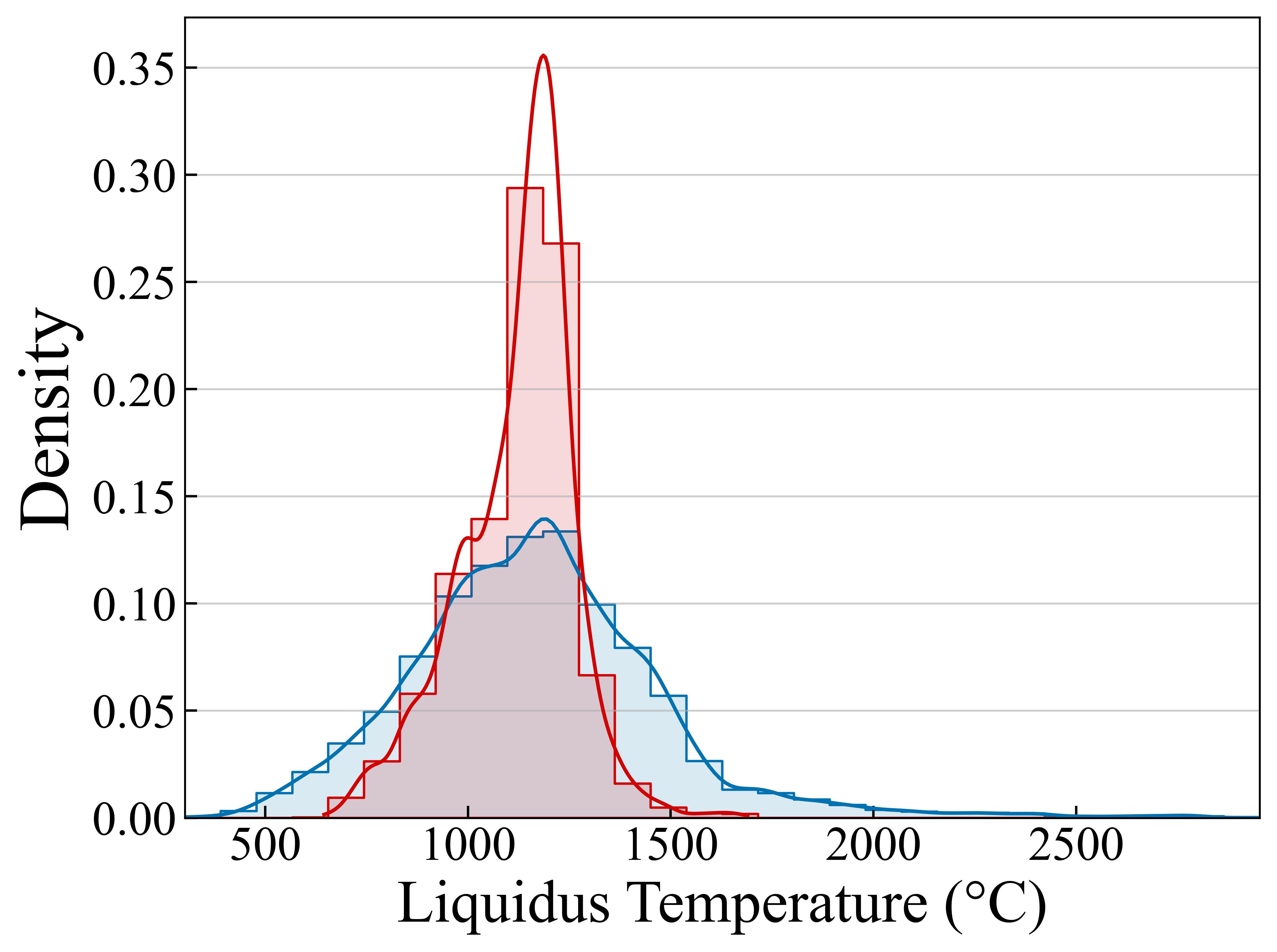}
\caption{}
\label{fig:tliqhist}
\end{subfigure}

\caption{Histograms of (a) refractive index, (b) Abbe number, and (c) liquidus temperature values.}
\label{fig:histograms}
\end{figure}

To evaluate the novelty of the collected compositions, histograms of the properties were constructed, where the x-axis represents the property values and the y-axis the frequency of occurrences. A density-based normalization was adopted instead of absolute counts, as comparing datasets of significantly different sizes is more informative when expressed in relative rather than absolute terms. Figures~\ref{fig:rihist},~\ref{fig:abbehist} and~\ref{fig:tliqhist} show, respectively, the histograms of refractive index, Abbe number and liquidus temperature.

From the analysis of the histograms, two main features can be highlighted. 
In panel~(a), the density of glasses with high refractive index values is greater than that observed in existing databases, indicating the presence of data that can effectively contribute to machine-learning models to predict and design new high-refractive-index glass systems.
In panel~(c), the curve exhibits a noticeable shift toward lower temperatures -- i.e., to the left -- along with a higher concentration of data around approximately~1000\,°C, suggesting that the extracted compositions predominantly belong to glass systems with relatively low liquidus temperatures.

\begin{figure}[H]
    \centering
    \includegraphics[width=0.7\linewidth]{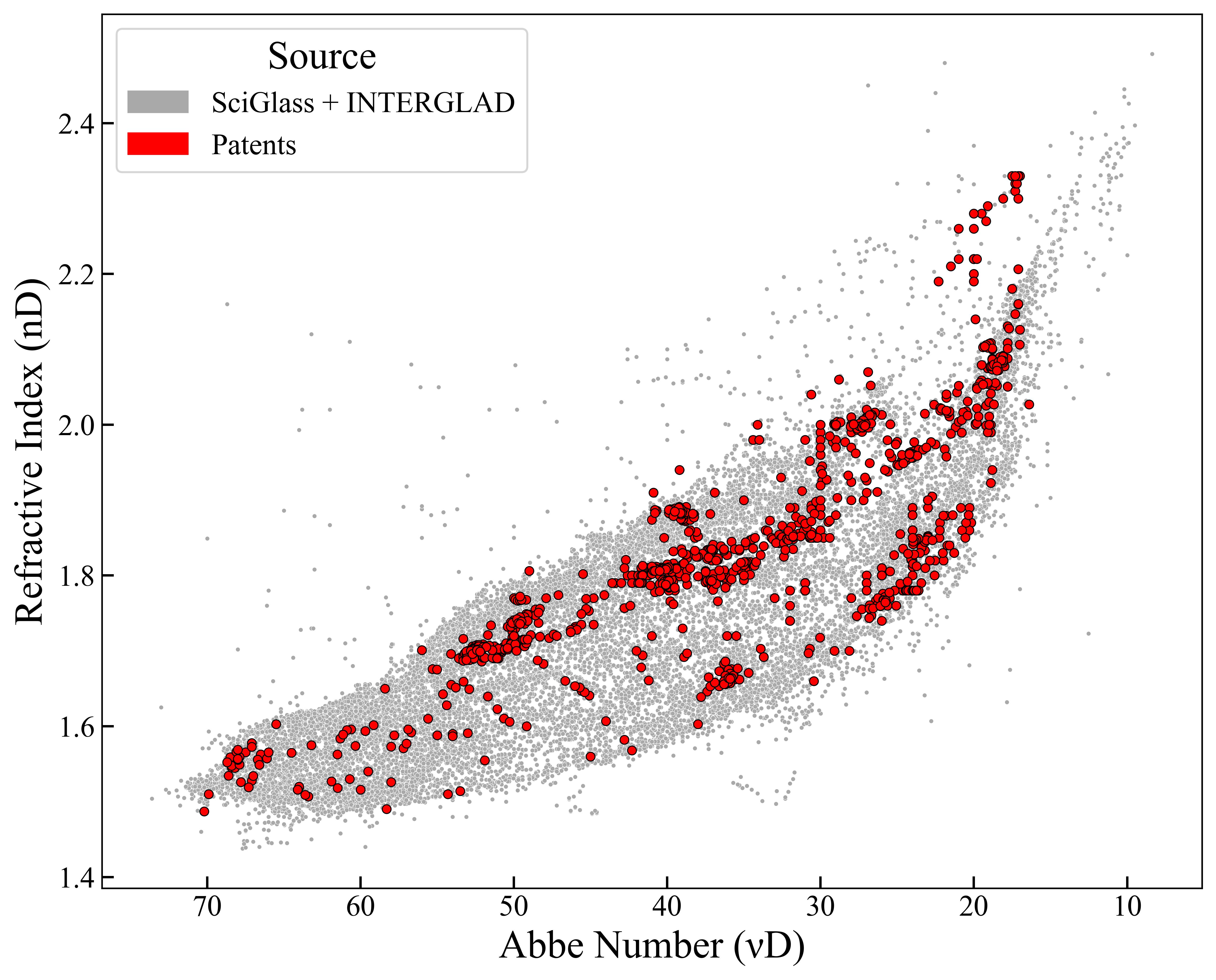}
    \caption{Abbe diagram with compositions from all three databases.}
    \label{fig:abbediagram}
\end{figure}

In further examining the novelty of the collected compositions, for the subset of data containing both refractive index and Abbe number measurements, a well-established representation in glass science -- known as the Abbe diagram -- was constructed (Figure~\ref{fig:abbediagram}). This diagram illustrates the correlation between the refractive index and the Abbe number for each composition. In this plot, the compositions extracted and compiled in the present work are highlighted in red, enabling a clear visual distinction between the newly retrieved data and those already reported in existing databases. This representation provides a more intuitive means to assess the extent to which the newly obtained patent data expand the compositional and optical property space of glass materials.

\begin{figure}[H]
    \centering
    \includegraphics[width=0.8\linewidth]{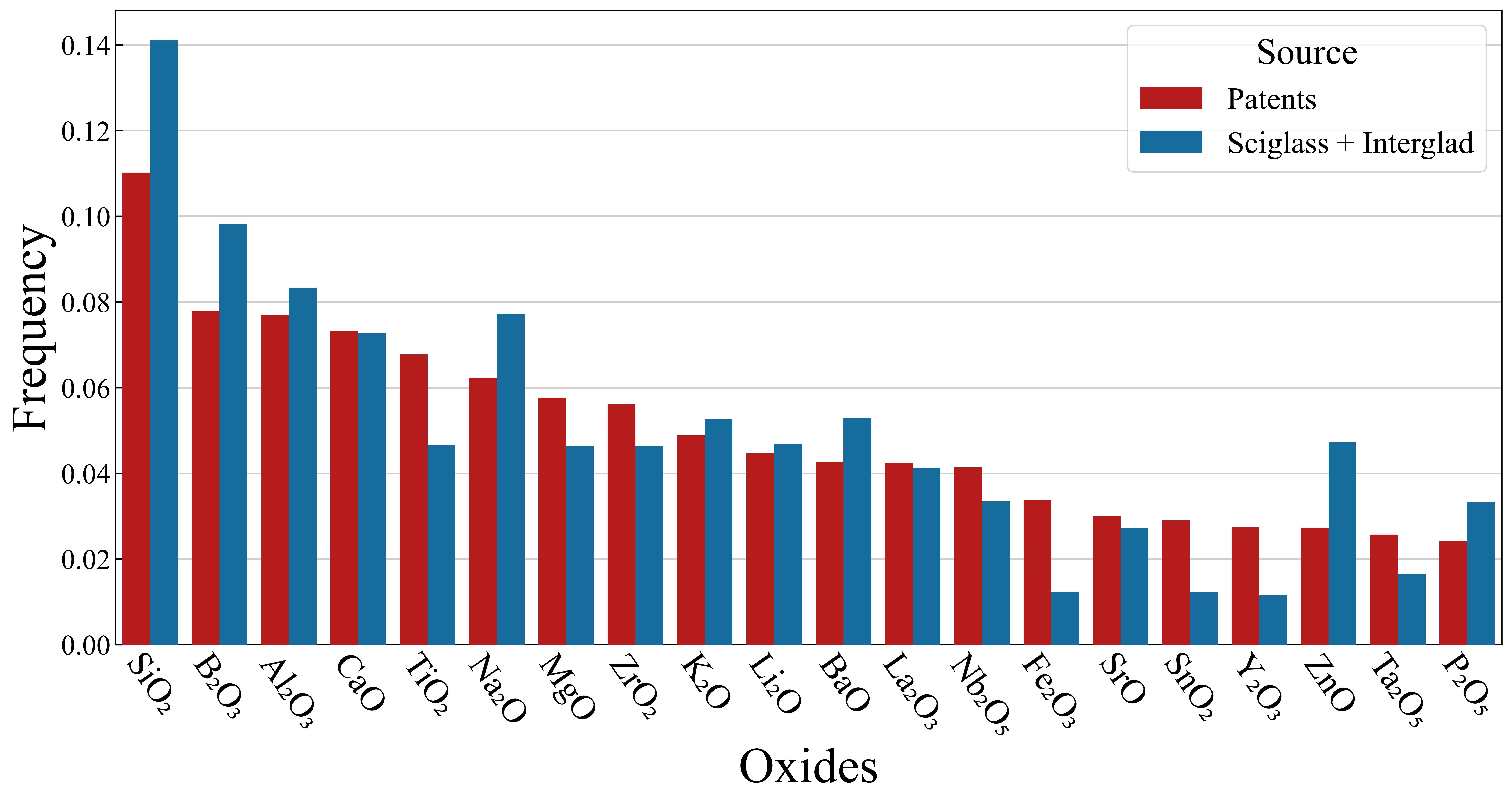}
    \caption{Relative frequency of the 20 most common oxides in the Patents database, with SciGlass+INTERGLAD comparison.}
    \label{fig:oxide_frequency}
\end{figure}

Beyond the analysis of optical properties, it is also instructive to investigate the compositional diversity of the dataset. To this end, the relative frequency of occurrence of each oxide component was quantified across all databases, defined as the number of occurrences of each oxide divided by the total number of oxide occurrences in the respective dataset. The oxides in the Patents dataset were then sorted in descending order according to their relative frequency within this dataset, enabling a consistent and direct comparison with the reference databases. For clarity of visualization, only the twenty most frequently occurring oxides were included in the plot. This analysis provides insight into the chemical breadth of the collected compositions and reveals which oxides dominate the compositional space of glass-forming systems. As shown in Figure~\ref{fig:oxide_frequency}, several oxides — particularly TiO$_2$, Fe$_2$O$_3$, SnO$_2$, and Y$_2$O$_3$ — exhibit a notably higher relative frequency of occurrence in the Patents dataset. Other oxides, such as MgO, ZrO$_2$, Nb$_2$O$_5$, SrO, and Ta$_2$O$_5$, display a moderately higher relative frequency compared to the reference databases. These observations reveal distinct compositional trends that diverge from those typically reported in Sciglass and INTERGLAD, further indicating that the Patents dataset contains novel features.

\begin{figure}[h]
    \centering
    \includegraphics[width=0.8\linewidth]{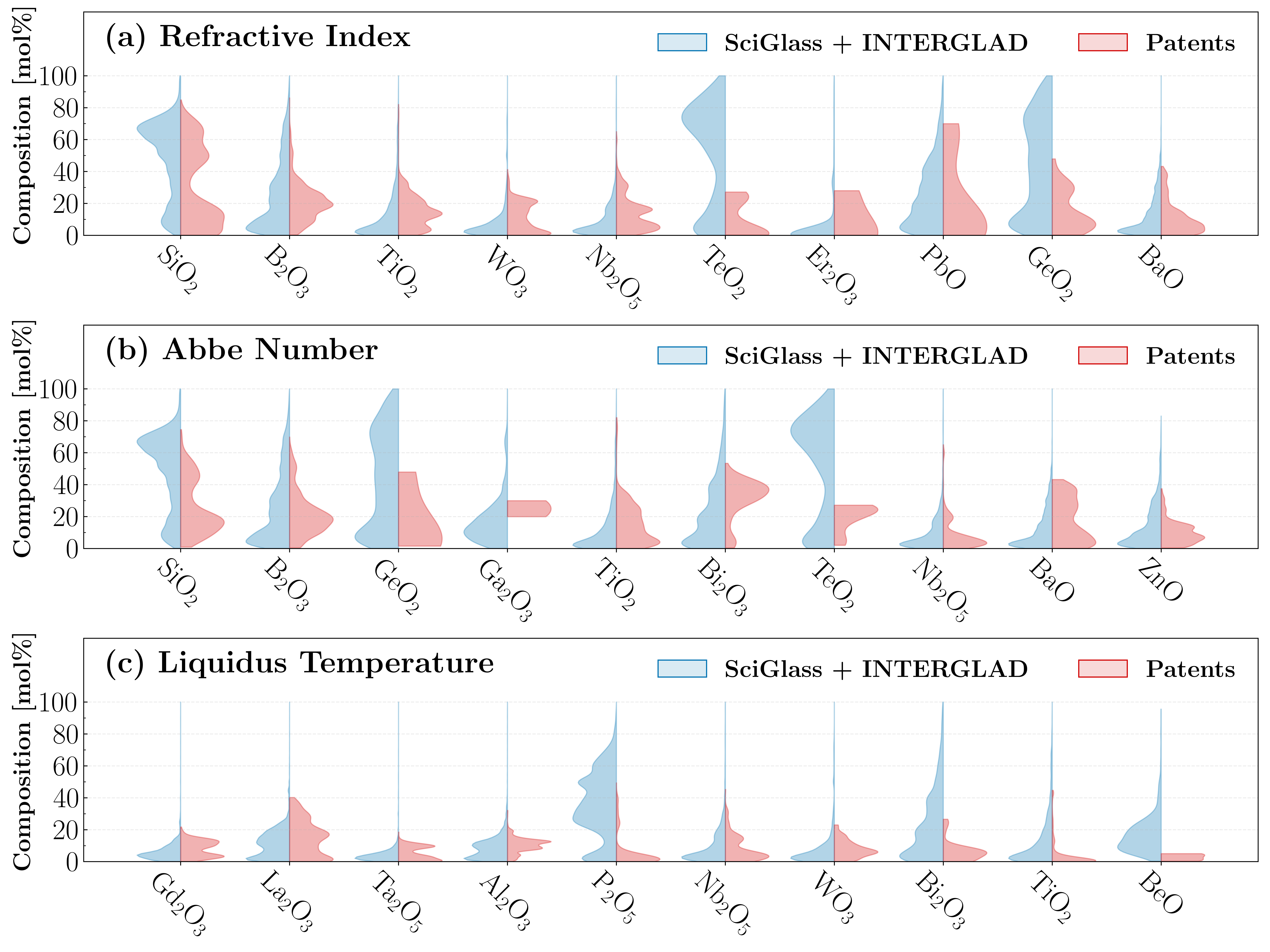}
        \caption{Violin plots illustrating the density distribution of selected oxides across the dataset for \ri, \abbe\ and \tliq.}
    \label{fig:violin}
\end{figure}

Moreover, the violin plots in Figure~\ref{fig:violin} illustrate the density distribution of selected oxides across the dataset. These plots highlight that we are introducing compositions in regions that are underrepresented in the combined SciGlass + INTERGLAD database. For instance, regarding Bi$_2$O$_3$ in the context of the Abbe number (Figure~\ref{fig:violin}b), we are adding compositions with more than 30\,mol\% of this oxide, a region that previously exhibited low density in SciGlass+INTERGLAD. This expansion of the compositional space can significantly impact the predictive performance of the models in regions that were previously sparsely populated.

\begin{figure}[h]
    \centering
    \includegraphics[width=\linewidth]{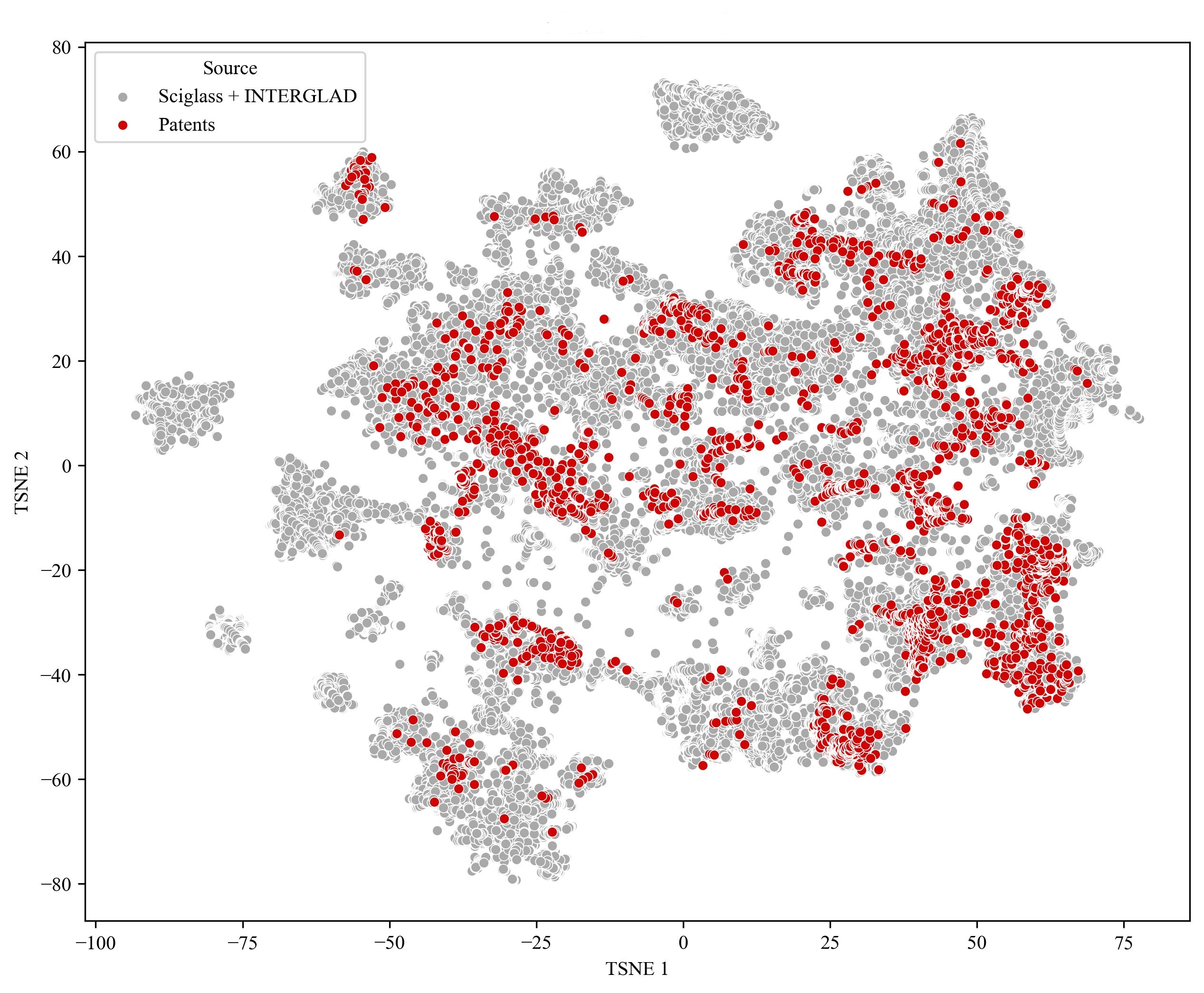}
\caption{t-SNE visualization of all glass compositions. Points in red correspond to compositions extracted from the Patents database.}
    \label{fig:tsne}
\end{figure}

Figure \ref{fig:tsne} provides an overview of how the compiled patent data are distributed within the existing compositional space. The glass compositions—which can contain up to 70 oxide components—were projected into two dimensions using t-SNE \citep{tsne_vandermaaten08a}. In this representation, points located close to one another correspond to compositions with similar chemical profiles, whereas distant points reflect more distinct compositions.


Table \ref{tab:patent_examples} illustrates types of unique glass compositions that the extraction pipeline is able to surface from patent literature but that are absent from INTERGLAD and SciGlass. Each entry represents a composition that occupies an extreme region of property space within the dataset -- low liquidus temperature for processability (US10106455B2\#7), exceptionally high Abbe number at low refractive index for chromatically stable optics (US20090122407A1\#13), and, conversely, very high refractive index combined with high dispersion for compact, high-power optical elements (US20240286947A1\#161). By bringing together composition, measured optical/thermal properties, and an application-oriented interpretation, the table shows that automated recovery of patent tables exposes technically meaningful, distinctive glass formulations that expand the design envelope available to data-driven glass discovery.

\begin{table}[H]
\centering
\caption{Examples of patent-extracted glass compositions [mol\%], their measured properties, and reported application significance.}
\label{tab:patent_examples}
\footnotesize
\renewcommand{\arraystretch}{1.25}
\setlength{\tabcolsep}{5pt}

\begin{tabularx}{\textwidth}{
>{\Centering\arraybackslash}m{3.5cm}   
>{\Centering\arraybackslash}m{2.4cm}   
>{\Centering\arraybackslash}m{1.8cm}   
>{\Centering\arraybackslash}m{1.8cm}   
>{\Centering\arraybackslash}m{1.8cm}   
>{\Centering\arraybackslash}m{3.0cm}   
}
\toprule
\textbf{Patent \# ID} & 
\makecell[c]{\textbf{Composition}\\{[mol\%]}} & 
\thead{\large $n$} &
\thead{\large $\nu_d$} &
\thead{$T_\text{liq}$\\{[°C]}} & 
\textbf{Application significance} \\
\midrule

\texttt{US10106455B2 Ex.\#7} &
\raisebox{0.6\height}{
\begin{tabular}[c]{@{}r@{ : }S[table-format=2.2]@{}}
Al$_2$O$_3$ &  3.23 \\
P$_2$O$_5$  & 41.75 \\
CaO         & 18.78 \\
MgO         &  8.16 \\
BaO         & 12.02 \\
K$_2$O      & 14.56 \\
CuO         &  1.51 \\
\end{tabular}
}&
Not measured & Not measured & 690 &
Near-infrared absorbing glass \\

\arrayrulecolor{lightgray}\midrule\arrayrulecolor{black}

\texttt{US20090122407A1 Ex.\#13} &
\raisebox{0.6\height}{
\begin{tabular}[c]{@{}r@{ : }S[table-format=2.2]@{}}
Al$_2$O$_3$ & 21.20 \\
P$_2$O$_5$  & 10.00 \\
B$_2$O$_3$  &  8.90 \\
CaO         & 30.50 \\
MgO         &  7.70 \\
SrO         & 21.80
\end{tabular}
}&
1.456 & 90.3 & Not measured &
Laser-writable, ultra-low-dispersion optical glass \\

\arrayrulecolor{lightgray}\midrule\arrayrulecolor{black}

\texttt{US20240286947A1 Ex.\#161} &
\raisebox{0.6\height}{
\begin{tabular}[c]{@{}r@{ : }S[table-format=2.2]@{}}
WO$_3$      & 27.00 \\
B$_2$O$_3$  & 15.00 \\
La$_2$O$_3$ & 17.00 \\
TiO$_2$     & 12.01 \\
Nb$_2$O$_5$ & 22.00 \\
ZrO$_2$     &  4.99 \\
Y$_2$O$_3$  &  2.00
\end{tabular}
}&
2.1583 & Not measured & Not measured &
High-refractive-index, rare-earth–containing optical glass \\
\bottomrule
\end{tabularx}
\end{table}

\section{Discussion and Future Work}\label{sec:discussion}

The present work represents an initial but essential step toward building a reliable, machine-learning-ready database of oxide glass compositions and properties directly extracted from patents. The current version of the pipeline successfully automates the retrieval, parsing, and cleaning of composition–property tables from Google Patents HTML pages, focusing on three key properties: refractive index (\ri), liquidus temperature (\tliq), and Abbe number (\abbe). 

The data extraction process generated information primarily from patents published in the last decade, a period during which the SciGlass database has not been updated. We observed a greater variety of compositions and properties values which should greatly enrich the fitting of predictive models and the development of new glasses. We emphasize that the compositions of the new database contain relatively more titanium, magnesium, zirconium, niobium, iron, tin, and yttrium oxides than those of the existing bases.

In future iterations, we aim to refine the data-cleaning phase by incorporating additional heuristics to improve consistency and facilitate property retrieval. In particular, enhanced rule-based filters and pattern-recognition methods will be designed to better identify measurement contexts, such as temperatures, wavelengths, molar masses, and other experimental parameters associated with property values. Moreover, large language models (LLMs) will be explored to assist in parsing semi-structured text segments and inferring missing metadata, improving the automatic recovery of property measurements and experimental conditions. 

We also plan to expand the scope of extracted properties beyond \ri, \abbe, and \tliq, including other thermomechanical, chemical, and optical attributes of interest for glass design -- such as viscosity, density, glass transition temperature, hardness, and chemical durability. Extending the property coverage will make the dataset a broader resource for diverse modeling applications in glass science.

At present, our scraping approach relies on parsing HTML structures from Google Patents. However, a significant fraction of patents provide only scanned PDFs or image-based tables. To address this limitation, upcoming work will integrate image-based extraction through table segmentation and optical character recognition (OCR). Combining these capabilities with the existing HTML-based pipeline will allow access to a much larger portion of the patent corpus.

Ultimately, this work contributes to a broader objective: enabling the discovery of oxide glass compositions with extreme or optimized properties through data-driven modeling. The patent-derived dataset developed here lays the foundation for future integration with machine learning methods aimed at predicting and inversely designing glasses with targeted performance across multiple dimensions.

\section*{Funding}
We are grateful to FAPESP -- The São Paulo Research Foundation -- for funding this work through a CEPID-CeRTEV project (process number: $2013/07793-6$) and four undergraduate research grants (processes number: $2025/01793-1$, $2025/01797-7$, $2025/02033-0$ and $2023/09945-0$).

\section*{Author Contributions} Conceptualization was performed by E.D.Z.; methodology by T.R.R., G.L.T, T.Y.B.A, E.T.C. and R.B.R.; formal analysis and investigation by G.L.T, T.Y.B.A and E.T.C.; writing -- original draft preparation -- by G.L.T, T.Y.B.A, T.R.R., D.A.Z and R.B.R.; writing -- review and editing -- by E.D.Z., T.R.R., D.A.Z., G.L.T, T.Y.B.A and R.B.R.; supervision by E.D.Z., T.R.R. and D.A.Z.

\bibliographystyle{apalike}
\bibliography{bibliography}

\end{document}